\documentclass{svproc}
\usepackage{graphicx}
\usepackage{url}

\newcommand{\sNN}{s_\mathrm{NN}}
\newcommand{\dd}{\partial}

\begin{document}
\mainmatter              
\title{Jet-fluid interaction in the EPOS3-Jet framework}
\titlerunning{Jet-fluid interaction in the EPOS3-Jet framework}  
%
\author{Iurii Karpenko\inst{1,2}\footnote{yu.karpenko@gmail.com} \and Joerg Aichelin\inst{1} \and
Pol Bernard Gossiaux\inst{1} \and \\ Martin Rohrmoser\inst{3,4} \and Klaus Werner\inst{1}}
\authorrunning{Iurii Karpenko et al.} 
%
\tocauthor{Iurii Karpenko, Joerg Aichelin, Pol Bernard Gossiaux, Martin Rohrmoser, Klaus Werner}
\institute{SUBATECH, Universit\'e de Nantes, IMT Atlantique, IN2P3/CNRS,\\ 4 rue Alfred Kastler, 44307 Nantes cedex 3, France,\\ \and
FNSPE, \v Cesk\' e vysok\' e u\v cen\' i technick\' e v Praze,\\ B\v rehov\'a 7, 11519 Prague 1, Czech Republic,\\ \and
Institute of Physics, Jan Kochanowski University, 25-406 Kielce, Poland,\\ \and
H.~Niewodnicza\'nski Institute of Nuclear Physics PAN, 31-342 Cracow, Poland
}

\maketitle              

\begin{abstract}
EPOS3-Jet is an integrated framework for jet modeling in heavy ion collisions, where the initial hard (jet) partons are produced along with soft (medium) partons in the initial state EPOS approach. The jet partons then propagate in the hydrodynamically expanding medium. The energy and momentum lost by the jet partons is added to the hydrodynamic medium 
via the source terms. The full evolution proceeds in a concurrent mode, without separating hydrodynamic and jet parts.
 
In this report we examine the medium recoil effects in Pb-Pb collisions at $\sqrt{\sNN}=2.76$~TeV LHC energy in the EPOS3-Jet framework.
\keywords{quark-gluon plasma, jets, relativistic hydrodynamics}
\end{abstract}
\section{Introduction}

A consistent modeling of back reaction of the hydrodynamic medium on the jet evolution is important for understanding the substructure of jets produced in heavy ion collisions. The majority of existing models implement only one-way jet-hydro interaction by coupling jets to a fixed hydrodynamic expansion and not including the energy deposition in the medium itself. On the other hand, some recent studies, e.g.\ \cite{Park:2018acg} conclude that the medium recoil effect has to be taken into account in order to reproduce the modification of the radial momentum distribution within a jet in AA collisions as compared to the $pp$ case. Therefore, in this report we examine the back reaction of the medium due to energy loss of jet partons in the EPOS3-Jet framework.

\section{Model}

The initial hard partons - seeds of the jets - are sampled from the initial state calculations in EPOS3 \cite{Werner:2013tya}. For the hydrodynamic expansion of the medium, the averaged initial conditions are taken. This makes it easier to visualize the effect of the jet energy loss on the medium.

Each initial hard parton leads to the development of a time-like parton cascade, due to collinear parton splitting caused by bremsstrahlung. The evolution of the parton cascade is performed with a Monte Carlo algorithm \cite{Rohrmoser:2018fkf} representing the Dokshitzer-Gribov-Lipatov-Altarelli-Parisi (DGLAP) equation with leading order $q\rightarrow qg$, $g\rightarrow gg$ and $g\rightarrow q\bar{q}$ splitting functions. The evolution of each parton cascade proceeds from an initial virtuality scale $Q_\uparrow$, which we set to be equal to parton's $p_\perp$, down to a minimal virtuality scale of $Q_\downarrow=0.6$~GeV.

The DGLAP evolution proceeds in momentum space. In order to couple the parton cascade to the medium, one has to make assumptions about its spacetime evolution. Therefore, we assume that in the global frame a parton has a mean life time (or the time before its next splitting occurs) of $\Delta t=E/Q^2$.

The parton shower has vacuum splitting functions. For the medium modifications of the parton shower,  we consider at present two effects: (i) an effective increase of the virtuality the off-mass-shell partons $\frac{dQ^2}{dt}=\hat{q}_R(T)$, where $t$ is time and $T$ the temperature in the local fluid (medium) rest frame. This mimics the medium induced radiation at the stage of jet formation; (ii) collisional energy loss which is modelled via a Langevin-type longitudinal drag and random transverse kicks to each jet parton:
$\Delta p_\parallel=-A(t,x) \Delta t, \quad
\Delta p_\perp = n_\perp \sqrt{\hat{q}_C \Delta t}.$
Different from the previously reported results \cite{Karpenko:2018uzf}, $A(t,x)$ and $\hat{q}_C$ are taken to be both temperature- and momentum-dependent according to \cite{Gossiaux:2009mk}, where they are evaluated with pQCD cross sections with running $\alpha_s$.

Here we are interested in a qualitative study of the jet-medium interaction and do not aim to fit the experimental data. Therefore, the transport coefficients $A(t,x)$ and $\hat{q}_C$ are further multiplied by a correction factor (or so-called K-factor) $K=0.166$ to approach the temperature-dependent jet transport coefficient from MARTINI model reported in \cite{Burke:2013yra}. This results in a parton-level $R_{\rm AA}\approx 0.6$ at $p_\perp^{\rm parton}=10...30$~GeV.\\
\textit{Back reaction on the medium.} The space-time evolution of the jet partons proceeds in the same time steps as the hydrodynamic evolution of the medium. During each time step, the energy and momentum loss of each jet parton in the computational frame of the fluid, is added to the corresponding fluid cells via an additional source term in the hydrodynamic equations: $\dd_{\nu} T^{\mu\nu}=J^\mu$. The source term $J^\mu$ is essentially a 4-vector of the lost energy and momentum in Milne coordinates, multiplied by a Gaussian smearing kernel with the width $R_g=0.4$~fm/c in the transverse direction around the position of the jet parton.

\begin{figure}
\includegraphics[width=0.9\textwidth]{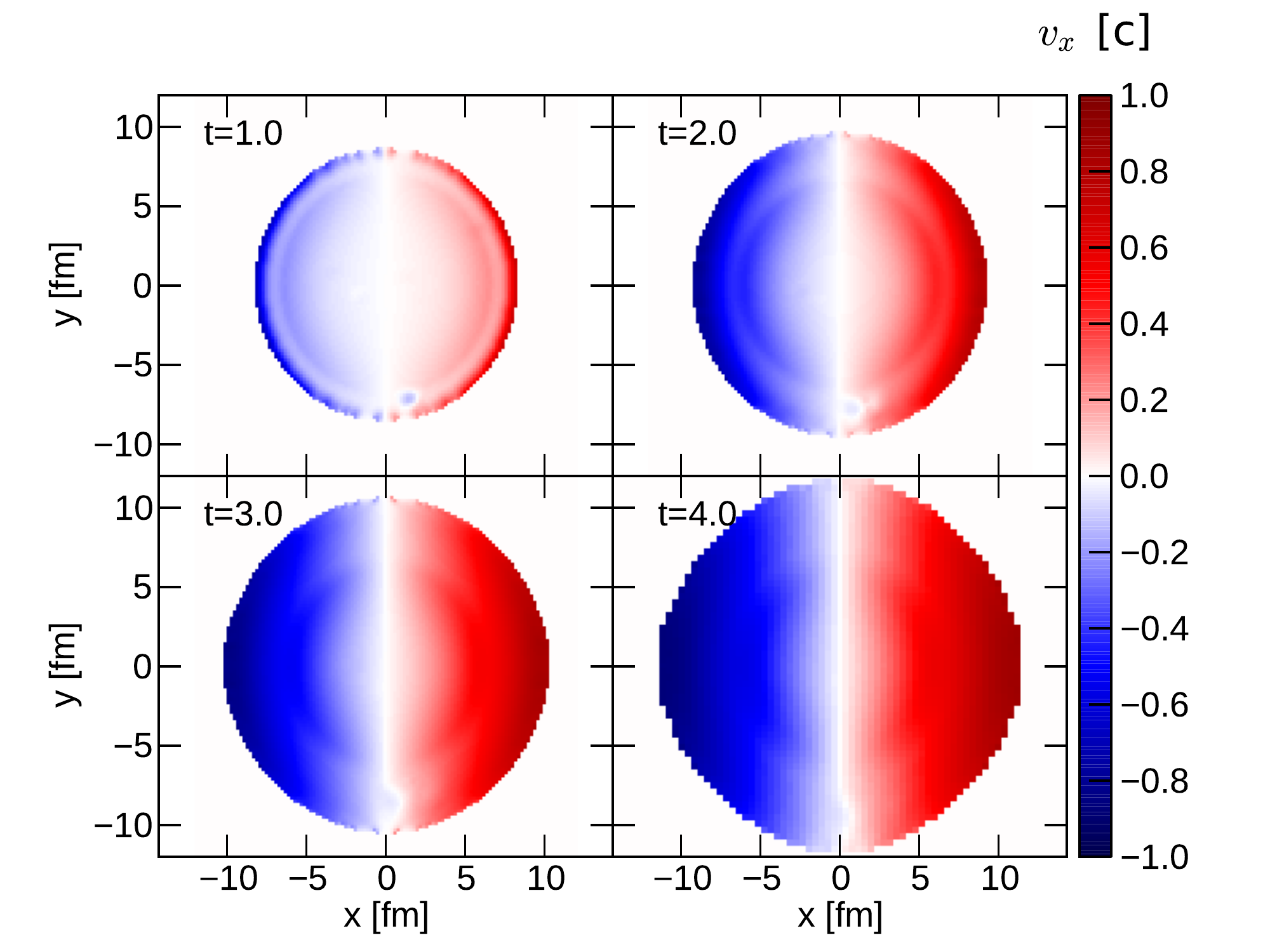}\\
\includegraphics[width=0.9\textwidth]{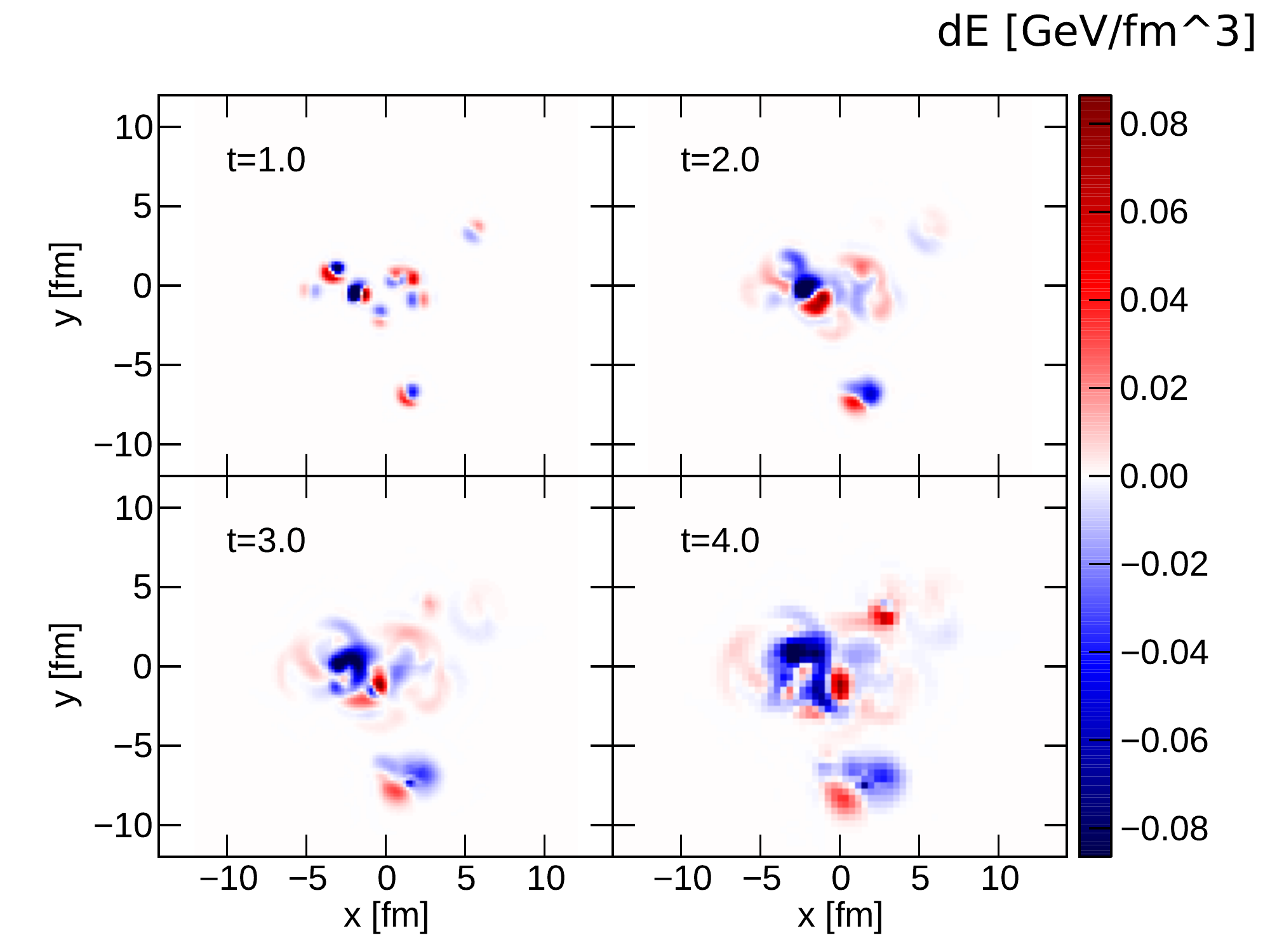}
\vspace*{-5pt}
\caption{$x$ component of transverse flow velocity (top panel) and corresponding perturbations in the local rest frame energy density of the medium caused by the energy lost by the jets (bottom panel). A jet event displayed here is simulated for 0-5\% central Pb-Pb collision at $\sqrt{\sNN}=2.76$~TeV.}\label{fig-machcones}
\end{figure}

\section{Results and conclusions}

The simulations are performed for most central Pb-Pb collisions at $\sqrt{\sNN}=2.76$~TeV collision energy.
To see the effects of medium recoil we run the ``semi-EbE'' simulations, where the sampled ensembles of initial hard partons are taken from the initial state EPOS with a requirement that in each event there must be at least 1 hard parton with $p_\perp>50$~GeV (so-called $p_\perp$ trigger), whereas the hydrodynamic evolution always starts from the smooth, event-averaged initial state. In other words, initial hard partons are sampled with Monte Carlo (EPOS) whereas the hydrodynamic initial state is not fluctuating.


In Fig.~\ref{fig-machcones} we show one of the events where the back reaction to the medium is visually expressed. On the top panel of the figure, which represents the snapshots of the $x$ component of the transverse flow velocity one can see an irregular spot which emerged already at $\tau=1$~fm/c. This local flip of collective flow velocity is caused by the energy loss of an energetic jet parton with $p_\perp\approx50$~GeV, which propagates through the medium in the $-x$ direction.

The bottom panel of Fig.~\ref{fig-machcones} shows perturbations in the local rest frame energy density of the fluid, i.e.\ a difference between a hydrodynamic evolution with the additional source term enabled and an unperturbed hydrodynamic evolution.
In the energy density perturbations, one can see more structures formed at $\tau=1$~fm/c, which all expand subsequently by a hydrodynamic evolution. This indicates that there are some 8 energetic jet partons at $\tau=1$~fm/c, which all lose their energy to the medium. One may recognize the ``Mach cones'' \cite{Betz:2008js} here, though the grid resolution does not permit to see them very clearly.

\textit{Conclusions.} Let us draw conclusions from Fig.~\ref{fig-machcones}. First, the \textit{relative} scale of energy density perturbations is very small in most of the cases. The perturbations become significant only at the periphery of the system where the background energy density is not large - which also results in visible velocity perturbations (top panel of Fig.~\ref{fig-machcones}). Second, the  hydrodynamic evolution smears out the early time perturbations, as one can see as well from the top panel of Fig.~\ref{fig-machcones}, and the expansion at late times becomes smooth again. Therefore we conclude that the present model for the jet-medium interaction with the assumption of instant thermalization of the energy-momentum loss by the jet partons makes it difficult to observe the medium recoil effects on the jet observables.

\textit{Acknowledgements}. The work is supported by Region Pays de la Loire (France) under contract no.~2015-08473.

%
%

\end{document}